\documentclass[12pt]{article}
\usepackage{a4,epsf}

\newcommand{\bea}{\begin{eqnarray}}
\newcommand{\eea}{\end{eqnarray}}
\newcommand{\simgt}{\hbox{ \raise3pt\hbox to 0pt{$>$}\raise-3pt\hbox{$\sim$} }}

\begin{document}
\begin{titlepage}
\title{
\vspace{2cm}
On the relation between QCD potentials in momentum
and position space}
\author{M.~Je\.zabek$^{a,b)}$,
M.~Peter$^{c)}$ and
Y.~Sumino$^{d)}$\thanks{On 
leave of absence from Department of Physics, Tohoku University,
Sendai 980-77, Japan.}
\\ \\ \\ \small
$a)$ Institute of Nuclear Physics,
Kawiory 26a, PL-30055 Cracow, Poland
\\   \small
  $b)$ Department of Field Theory and Particle Physics, University of 
      Silesia, \\[-5pt]  \small
     Uniwersytecka 4, PL-40007 Katowice, Poland\\ \small
$c)$ Institut f\"ur Theoretische Physik,
Universit\"at Heidelberg, \\ [-5pt]
\small
Philosophenweg 16, D-69120 Heidelberg, Germany
\\
\small
$d)$ Institut f\"ur Theoretische Teilchenphysik,
Universit\"at Karlsruhe,\\  [-5pt]
\small
D-76128 Karlsruhe, Germany
}
\maketitle
\thispagestyle{empty}
\vspace{-5.5truein}
\begin{flushright}
{\bf HD-THEP--98--10}\\
{\bf TTP 98-09}\\
{\bf TP-USl/98/3}\\
{\bf hep-ph/9803337}\\
{\bf March 1998}
\end{flushright}
\vspace{4.0truein}
\begin{abstract}
We derive a formula which relates the QCD potentials in
momentum space and in position space in terms of
the $\beta$ function of the renormalization-group
equation for the potential.
This formula is used to study the theoretical uncertainties
in the potential and in particular in its application to the
determination of the pole mass $m_b$ when we use perturbative expansions.
We demonstrate the existence of these uncertainties for the Richardson
potential explicitly and  then discuss
the limited theoretical accuracy in the perturbative QCD potential.
We conclude that a theoretical uncertainty of $m_b$ 
much below 100~MeV would not be achievable within perturbative QCD.
\\[1\baselineskip]
\end{abstract}
\end{titlepage}
  
In this article we discuss a relation between the static QCD 
potential~\cite{statpot}
in momentum space
\bea
V(q) = - C_F \, \frac{4\pi\alpha_{_V}(q)}{q^2} ,
\label{mompot}
\eea
where for quarks $C_F=4/3$, and the corresponding potential in position
space\footnote{We use the notation $q=|{\bf q}|$ and $r=|{\bf r}|$.}
\bea
\bar{V}(r) = \int \frac{d^3{\bf q}}{(2\pi)^3} \,
V(q) \, e^{i{\bf q} \cdot  {\bf r}}
= - C_F \frac{\bar{\alpha}_{_V}(1/r)}{r} .
\label{coordpot}
\eea
Eqs.~(\ref{mompot}) and (\ref{coordpot}) 
are understood as the definitions of the couplings $\alpha_{_V}(q)$
and $\bar{\alpha}_{_V}(1/r)$.
In addition, they express the fact that these two functions are related
because $\bar{V}(r)$ is the Fourier transform of $V(q)$.
We can obtain a useful expression for this functional relation
\bea
\bar{\alpha}_{_V}(1/r) = F[ \alpha_{_V}(q)] 
\label{functional}
\eea
using the renormalization-group equation of the coupling $\alpha_{_V}(q)$,
and we will study the consequences of this relation.

First we relate the coupling at a general scale, $\alpha_{_V}(q)$,
to its value at a specific 
renormalization scale $\mu$.
Let us express $\alpha_{_V}(q)$ as a power series
\bea
\alpha_{_V}(q) \equiv \Phi ( \alpha_{_V}(\mu),t) =
\sum_{n=0}^\infty \,  c_n (\mu) \, t^n \qquad
\mbox{where}\quad t = \ln ( \mu^2/q^2 ).
\label{Phi}
\eea
It follows that 
\bea
\alpha_{_V}(\mu) = \Phi( \alpha_{_V}(\mu),0) = c_0(\mu).
\label{boundarycond}
\eea
For static heavy quarks, the potential energy 
is a physical quantity.
As a result, $\alpha_{_V}(q)$ 
obeys the renormalization-group equation
\bea
\mu^2 \frac{d}{d\mu^2} \, \alpha_{_V}(q) =
\frac{\partial \Phi ( \alpha_{_V},t)}{\partial t} +
\beta_{_V}(\alpha_{_V}) \,
\frac{\partial \Phi ( \alpha_{_V},t)}{\partial \alpha_{_V}} = 0,
\eea
where the $\beta$ function for $\alpha_{_V}$ is defined by
\bea
\beta_{_V}(\alpha_{_V}) \equiv
\mu^2 \frac{\partial \alpha_{_V}(\mu)}{\partial \mu^2} 
= - 4\pi \sum_{n=0}^\infty
\beta_{_{V,n}}
\left(
\frac{\alpha_{_V}(\mu)}{4\pi}
\right)^{n+2} .
\label{betafn}
\eea
Since the coefficients $c_n(\mu)$ in eq.~(\ref{Phi}) can be 
expressed as partial 
derivatives of $\Phi$ with respect to $t$,
\bea
c_n(\mu) = \frac{1}{n!}
\left. \frac{\partial^n}{\partial t^n} \, \Phi ( \alpha_{_V}(\mu),t)
\right|_{t=0} ,
\label{cnmu}
\eea
one can show using the renormalization-group equation
\bea
c_n (\mu ) = \frac{1}{n!} \left(
- \beta_{_V} (\alpha_{_V}) \, \frac{\partial}{\partial \alpha_{_V}} 
\right)^n
\Phi (\alpha_{_V},t=0)
= \frac{1}{n!}
\left( - \beta_{_V} (\alpha_{_V}) \, \frac{\partial}{\partial \alpha_{_V}} 
\right)^n
\alpha_{_V}(\mu) .
\eea
Combining this expression with eqs.~(\ref{Phi}) and 
(\ref{cnmu}), we find the relation between the
couplings $\alpha_{_V}(q)$ and $\alpha_{_V}(\mu)$:
\bea
\alpha_{_V}(q) = 
\exp \left[
{
- t \, \beta_{_V} (\alpha_{_V}) \, \frac{\partial}{\partial \alpha_{_V}} 
} \right] \, \alpha_{_V}(\mu) =
\sum_{n=0}^\infty \,
\frac{t^n}{n!} 
\left(
- \beta_{_V} (\alpha_{_V}) \, \frac{\partial}{\partial \alpha_{_V}} 
\right)^n \alpha_{_V}(\mu) .
\label{rgevolution}
\eea
From this expression, one finds
\bea
c_n (\mu) = {\cal O}\left(\, [ \alpha_{_V}(\mu)]^{n+1}\, \right) .
\eea

Now we can perform the Fourier transform in eq.~(\ref{coordpot})
using the following formula:
\bea
\int \frac{d^3{\bf q}}{(2\pi)^3} \, \,
t^n \, \frac{e^{i{\bf q} \cdot  {\bf r}}}{q^2}
= \left.
\frac{\partial^n}{\partial u^n} \,
{\cal F}(r,\mu,u)
\right|_{u=0} ,
\label{fourierlog}
\eea
where for $-1<u<1/2$,
\bea
{\cal F}(r,\mu,u) \equiv \mu^{2u}
\int \frac{d^3{\bf q}}{(2\pi)^3} \,
\frac{e^{i{\bf q}\! \cdot \! {\bf r}}}{(q^2)^{1+u}}
=
\frac{(\mu r^\prime )^{2u}}{4\pi r} \, f(u) ,
\label{f}
\eea
with
\bea
f(u) &=& \sqrt{\frac{\tan (\pi u)}{\pi u}} \,
\exp \left[
\sum_{n=1}^\infty \frac{(2u)^{2n+1}}{2n+1} \zeta(2n+1) 
\right]
\nonumber \\  &=&
\sum_{n=0}^\infty f_n \, u^n
\nonumber \\  &=&
1 + {{{{\pi }^2}\,{u^2}}\over 6} + 
    {{8\,{\zeta}(3)\,{u^3}}\over 3} + 
   {{19\,{{\pi }^4}\,{u^4}}\over {360}} + 
   \left( {{4\,{{\pi }^2}\,{\zeta}(3)}\over 9} + 
      {{32\,{\zeta}(5)}\over 5} \right) \,{u^5} + \cdots
\label{expf}
\eea
Here,
\bea
r^\prime = r \, e^{\gamma_{_E}} ,
\eea
$\gamma_{_E} = 0.57721...$ is Euler's constant,
and $\zeta$ denotes the Riemann $\zeta$ function.\footnote{
Different expressions for ${\cal F}(r,\mu,u)$ are derived in~\cite{mp}.
One of those,
\bea
{\cal F}(r,\mu,u) = \frac{(\mu r)^{2u}}{4\pi^2r} \,
\frac{\Gamma( \frac{1}{2}+u) \, \Gamma( \frac{1}{2}-u) }
{\Gamma(1+2u)} ,
\nonumber
\eea
can be used to derive eqs.~(\ref{f}) and (\ref{expf})
via the formulas \cite{c4}
\bea
\Gamma( \mbox{$\frac{1}{2}$}+z) \, \Gamma( \mbox{$\frac{1}{2}$}-z)
= \frac{\pi}{\cos (\pi z)}
\nonumber
\eea
and 
\bea
\ln \, \Gamma(1+z) = \frac{1}{2}\, \ln \,
\frac{\pi z}{\sin (\pi z)} - \gamma_E z -
\sum_{n=1}^\infty \frac{z^{2n+1}}{2n+1} \, \zeta (2n+1) .
\nonumber
\eea
}
Higher order terms in the expansion of $f(u)$ at $u=0$ can be
easily obtained using e.g.\ Mathematica \cite{mathematica}.
It follows from eqs.~(\ref{fourierlog}) and (\ref{f}) that
\bea
F[t^n] = \left. 
\frac{\partial^n}{\partial u^n} \,
[ e^{uy} \, f(u) ]  \right|_{u=0}
= \sum_{k=0}^n \left( 
\begin{array}{c}
n \\ k
\end{array} \right)
\, y^k \,
\frac{\partial^{n-k} f(u)}{\partial u^{n-k}} \biggl|_{u=0}
= \sum_{k=0}^n \frac{n!}{k!} \, f_{n-k} \, y^k ,
\eea
where 
\bea
y = \ln (\mu^2 {r^\prime}^2 ) .
\eea
Then from eq.~(\ref{rgevolution}) we obtain
\bea
\bar{\alpha}_{_V} (1/r) = F[\alpha_{_V}(q)]\,  
=\, \sum_{n=0}^\infty \frac{F [t^n] }{n!}
\left(
- \beta_{_V} (\alpha_{_V}) \, \frac{\partial}{\partial \alpha_{_V}} 
\right)^n \alpha_{_V}(\mu) .
\label{rgevolution2}
\eea
In particular, for $y=0$, which implies $\mu = 1/r^\prime$,
\bea
F[t^n](y=0) = n! \, f_n .
\eea
Hence,
\bea
\bar{\alpha}_{_V}(1/r) =
\sum_{n=0}^\infty f_n
\left(
- \beta_{_V} (\alpha_{_V}) \, \frac{\partial}{\partial \alpha_{_V}} 
\right)^n
\alpha_{_V}(q= 1/r^\prime) .
\label{result}
\eea
This equation is the specific representation of the functional relation 
(\ref{functional}) which we set out 
to derive. The function $\beta_{_V}(\alpha_{_V})$ is defined in
eq.~(\ref{betafn}) and the coefficients $f_n$ in eq.~(\ref{expf}).

\begin{figure}[t]\begin{center}
\vspace{2cm}
 \epsfxsize 12cm \mbox{\epsfbox{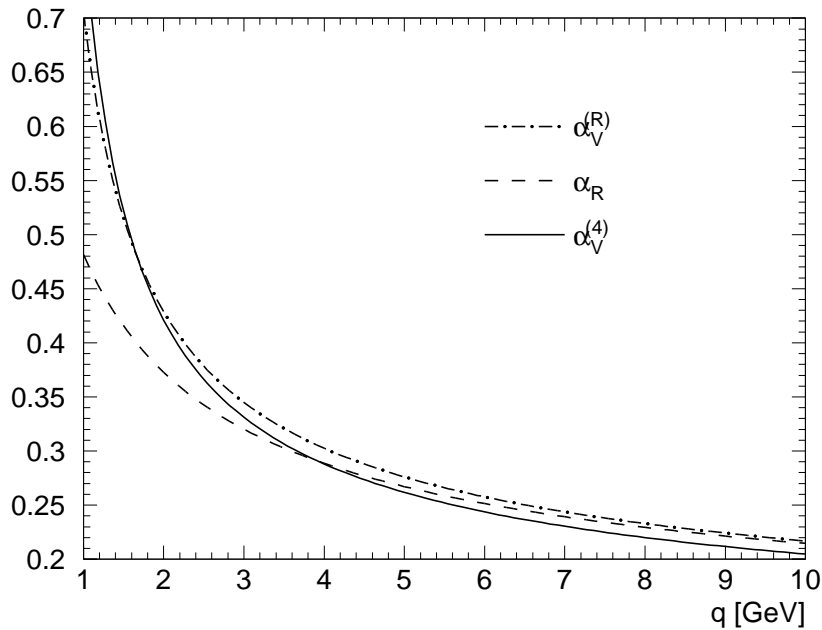}}
 \end{center}
 \caption[]{\label{fig1}Comparison of the 
momentum-space couplings: $\alpha_{_V}^{(4)}(q)$ --
solid, $\alpha_{_V}^{({\rm R})}(q)$ -- dash--dotted, 
and $\alpha_{_{\rm R}}(q)$ --
dashed line.  }
\end{figure}

Before studying consequences of eq.~(\ref{result}) for the 
perturbative QCD potential,
let us discuss first a model which we can solve numerically and which
describes the energy levels of quarkonia very well.
The model that we consider is the Richardson potential~\cite{richardson}
which in momentum space reads:
\bea
V_{\rm R} (q)\, =\,- 4\pi C_F \frac{\alpha_{_V}^{(R)}(q)}{q^2}
\, =\, - \frac{16\pi^2\, C_F}{\beta_0\, q^2 
\ln ( 1 + q^2/\Lambda_{\rm R}^2)} \; \cdot
\eea
A very good description of the charmonium and bottomonium states is 
obtained using this potential for $\Lambda_{\rm R} = 0.4$~GeV and
\bea
\beta_0 = 11 - \frac{2}{3}n_f ,
\eea
where $n_f = 3$ is the number of the light quarks. In Fig.\ref{fig1}
the function $\alpha_{_V}^{(\rm R)}(q)$ is plotted as the dash--dotted line
for $q$ between 1 and 10~GeV. 
The next-to-next-to-leading order 
(three-loop renormalization-group improved) QCD coupling 
$\alpha_{_V}^{(4)}(q)$~\cite{mp} 
for $n_f=4$ active flavours and $\alpha_{\overline{MS}}(m_Z)$= 0.119 
is also shown as the solid 
line.\footnote{
We used the three-loop renormalization-group equation 
to evolve the $\overline{MS}$ coupling, and Eq.~(9) of~\cite{chetyrkin}
without the $a^3$-term to match the
4-flavour to the 5-flavour theory at the matching scale 5~GeV,
and $m_b({\rm pole})=4.88$~GeV.
The result is not very sensitive to varying this matching scale.
}
It is seen that in the range of $q$
relevant for the position of $\Upsilon(1S)$ the two curves are fairly
close. The QCD potential is more attractive which implies that the
value of $m_b$ extracted from the mass of $\Upsilon(1S)$ is slightly
larger for the next-to-next-to-leading order QCD than for 
the Richardson potential. 
In fact in a recent article~\cite{PY} 
$m_b= 4.96$~GeV is obtained\footnote{
This value is different from what~\cite{PY} gives as a final value
of $m_b$.
It corresponds to the result from the two-loop static QCD potential,
which does not include a shift in $m_b$ of about 50~MeV due to 
relativistic and leading non-perturbative corrections.
} 
with an uncertainty
of about 100~MeV arising from the uncertainty in the input value of 
$\alpha_{\overline{MS}}(m_Z)$. For the Richardson potential $m_b= 4.88$~GeV
is obtained.
All these attractive features suggest 
that at scales of a few GeV the model closely 
resembles the true QCD potential.

For this model, we study the validity of the relation (\ref{result}) 
when it is expressed as a perturbative series of $\alpha_{_V}$
in the asymptotic region where the coupling is small.
In order to define the perturbative coupling of this model unambiguously
in both momentum and position spaces,
we subtract the confining part of the Richardson
potential 
\bea
V_{\rm R} (q) = V_{\rm conf}(q) - 4\pi C_F \frac{\alpha_{_{\rm R}}(q)}{q^2} .
\eea
The confining part $V_{\rm conf}(q)$ 
corresponds to a linear confining potential in position space
\bea
\bar{V}_{\rm conf}(r) = \frac{2\pi C_F}{\beta_0} \Lambda_{\rm R}^2 r .
\label{linearpot}
\eea
The coupling
\bea
\alpha_{_{\rm R}}(q) = 
\frac{4\pi}{\beta_0}
\left[
\frac{1}{\ln ( 1 + q^2/\Lambda_{\rm R}^2)} -
\frac{\Lambda_{\rm R}^2}{q^2}
\right] 
\label{alphaR}
\eea
is considered as a perturbative coupling in momentum space.
The coupling $\alpha_{_{\rm R}}(q)$ is plotted as the dashed line
in Fig.\ref{fig1}.
The corresponding coupling in position space is given by 
the following expression~\cite{richardson}:
\bea
\bar{\alpha}_{\rm R}(1/r) =
\frac{2\pi}{\beta_0}
\left[ 1 - 4 \int^\infty_1 \frac{du}{u}
\frac{e^{-\Lambda_{\rm R}ru}}{\ln^2 ( u^2 - 1) + \pi^2}
\right] 
\label{richcoupling}
\eea
which can be calculated numerically to a very high precision.
In particular the accuracy of our numerical calculations was $10^{-10}$.
From eq.~(\ref{alphaR}) we calculate the $\beta$ function
\bea
\beta_{_{\rm R}}(\alpha_{_{\rm R}}) = q^2 
\frac{\partial \alpha_{_{\rm R}}(q)}{\partial q^2} =
- \, \frac{\beta_0}{4\pi} \alpha_{_{\rm R}}^2 +
\frac{\beta_0}{4\pi} \,
\frac{ [ \frac{4\pi}{\beta_0}-\alpha_{_{\rm R}} ]^2}{1+q^2/\Lambda_{\rm R}^2}
,
\eea
where $q^2/\Lambda_{\rm R}^2$ is understood as a function of
$\alpha_{_{\rm R}}$.
For $q \gg \Lambda_{\rm R}$ the second term in the expression for
$\beta_{\rm R}$ is negligible.
Then our problem becomes very simple: there is only one term in the
sum in eq.~(\ref{betafn}) corresponding to
\bea
\beta_{{\rm R},0} = \beta_0 .
\eea
Instead of eq.~(\ref{result}) we have a much simpler relation
\bea
\bar{\alpha}_{\rm R}(1/r) \sim
\sum_{n=0}^\infty \alpha_{_{\rm R}}(q)
\left(
\frac{\beta_0\, \alpha_{_{\rm R}}(q)}{4\pi}
\right)^n n! \, f_n ,
\label{relrich}
\eea
with $q=1/r^\prime$.
The symbol ``$\sim$'' indicates that in our derivation non-perturbative 
higher twist contributions to $\beta_{\rm R}$ 
(terms suppressed by powers of $e^{-4\pi/( \beta_0 \alpha_{\rm R})}$)
have been neglected.
Strictly speaking, the relation (\ref{relrich}) may be valid only for
$q \gg \Lambda_{\rm R}$ i.e.\ for small $\alpha_{_{\rm R}}$.
We have computed the first twelve coefficients $f_n$ and checked that all
are positive.
For a given $1/r^\prime$, the terms in (\ref{relrich}) decrease with 
increasing
$n$ for $n < n_{\rm min}$ and increase for $n > n_{\rm min}$.
The value $n_{\rm min}$ corresponding to the minimal contribution 
increases with growing $1/r^\prime$.
The best approximation is obtained when the series in (\ref{relrich})
is truncated at $N \simeq n_{\rm min}$.
\begin{table}
\begin{center}
\begin{tabular}{|c|r|l|l|c|c|} \hline  
      &      &               &            &           & \\
$1/r$ [GeV]      &   N   &  $\  \bar\alpha_{_{R,N}}(1/r)$   
&\ \  $\bar\alpha_{_{R}}(1/r)$   &
$\delta\bar V_-$ [GeV]  &  $\delta\bar V_+$ [GeV] \\ 
      &      &               &            &           & \\  \hline 
$10$  &  3   &  0.2856       &  0.2714    &  -0.193   &  0.189    \\  \hline
$20$  &  3   &  0.22262      &  0.22183   &  -0.381   &  0.021    \\  \hline
$50$  &  4   &  0.17647      &  0.17470   &  -0.222   &  0.118    \\  \hline
$10^2$&  4   &  0.14868      &  0.14912   &  -0.059   &  0.183    \\  \hline
$10^3$&  6   &  0.098937     &  0.099017  &  -0.106   &  0.120    \\  \hline
$10^4$&  9   &  0.07413536   &  0.07413084 &  -0.143   &  0.060    \\  \hline
$10^5$& 11   &  0.05938222   &  0.05938213 &  -0.175   &  0.012    \\  \hline
$10^6$& 12   &  0.04958339   &  0.04958315 &  -0.409   &  0.318    \\  \hline
$10^7$& 12   &  0.04257994   &  0.04257992 &  -1.008   &  0.185    \\  
\hline
\end{tabular}
\end{center}
\caption{Comparison of the coupling $\bar\alpha_{_{R}}(1/r)$
for the Richardson potential and the best approximation obtained
by truncating the asymptotic series (\protect\ref{relrich})
at $n=N$. }
\label{table1}
\end{table}
In Table~\ref{table1} the values are given of the functions
$\bar{\alpha}_{\rm R}(1/r)$ computed numerically from 
eq.~(\ref{richcoupling}) and $\bar{\alpha}_{\rm R,N}(1/r)$
obtained from the asymptotic expansion (\ref{relrich})
truncated at $n = N$.
The coefficient $\beta_0$ is evaluated for $n_f=3$
and $1/r$ is varied between $10$ and $10^7$~GeV.
It is evident that for large values of $1/r$ a very good approximation is
obtained using the truncated series $\bar{\alpha}_{\rm R,N}(1/r)$.
However, at $1/r = 10$~GeV, $n_{\rm min}=1$ and the quality of approximation
practically does not improve when instead of
\bea
\bar{\alpha}_{_{\rm R}}(1/r) \simeq \alpha_{_{\rm R}}(q=1/r^\prime)
\eea
the formula including the ${\cal O}(\alpha_{_{\rm R}}^3)$ term is used.
At $1/r = 20$~GeV inclusion of the cubic term improves the quality of
approximation, and at $1/r = 100$~GeV also the term 
${\cal O}(\alpha_{_{\rm R}}^4)$ is needed.
Let the truncution at $n=n_+$ result in the closest approximation of
$\bar{\alpha}_{\rm R}$ from above  and $\delta\bar V_+(r)$ denote the
difference between the exact and the approximate
values of the potential $\bar V_{\rm R}(r)$.
If the contribution of the term for $n=n_+$ is not included,
i.e. the series in (\ref{relrich}) is truncated at  $n=n_+ -1$ 
the closest approximation from below is obtained.
Let $\delta\bar V_-(r)$ be the
corresponding difference of the potentials. The values of the functions
$\delta\bar V_\pm(r)$ are also given in Table~\ref{table1}.
One can estimate the theoretical uncertainty of the perturbative
approach by considering the change in the value of the potential
\bea
\delta\bar V(r) = \delta \overline{V}_+ - \delta \overline{V}_-  
\eea
corresponding to truncating the asymptotic formula at $n=n_+ -1$
and $n=n_+$.
It is interesting\footnote{
This can be shown easily for a small $\alpha_{\rm R}$:
Since asymptotically $f_n \sim 2^n$, it follows that
$n_{\rm min} \sim 2\pi /( \beta_0 \alpha_{\rm R} ) 
\sim \ln (1/(\Lambda_{\rm R}r))$.
Then one can estimate the size of the last term included or rejected
in eq.~(\ref{relrich}) to be 
$\delta \bar{\alpha}_{\rm R} \sim \Lambda_{\rm R} r$, which leads to
$\delta \bar{V}(r) \sim \Lambda_{\rm R}$.
}
that $\delta\bar V(r)$ does not change drastically with 
$r$ and is of the order of $\Lambda_{\rm R}$.

In perturbative calculations of the energy levels of a $Q\bar{Q}$ 
system in this model, the truncation of the asymptotic series for the potential
leads to an uncertainty
in the perturbative pole mass $m_Q$ of the heavy quark.
This uncertainty is of the order of
$\delta m \sim \delta\bar V(r_0)/2$, where $r_0$ denotes the typical size 
of the bound-states. This estimation is obtained assuming that the
shift in the binding energy due to the change of the coupling 
$\bar{\alpha}_{_{\rm R}}$
is equal to $\;-\; \delta\bar V(r_0)$, which would be the case for the
Coulomb potential. The shift in the binding energy is then  
compensated by shifting the masses of $Q$ and $\bar Q$ by $\delta m$. 
One may think that for a stable quark $Q$ the uncertainty in $m_Q$
can be reduced by comparing the
$Q\bar Q$ bound states of different sizes, thereby disentangling
the correlation in the dependences on the coupling and mass.
However, as it has already been mentioned the variation of
$\delta\bar V(r)$ with $r$ is quite moderate. 
Moreover at larger
distances non-perturbative effects become important.
So, in fact it is difficult to reduce the uncertainty in the mass
determination significantly.\footnote{
Behind this argument for the Richardson potential
lies a following corresponding perspective in QCD.
In principle one is able to determine
the two independent fundamental parameters
of QCD, $m_Q$ and $\alpha_s$, from
a sufficient number of different physical observables 
${\cal O}_i(m_Q,\alpha_s)$ by exploiting their
different dependences on these parameters.
When we express the observables in perturbative expansions
of $\alpha_s$, however, we encounter uncertainties of the
order of $\Lambda_{QCD} \sim 200$~MeV originating from truncations of 
the asymptotic series.
Then, if we try to extract the values of $m_Q$ and $\alpha_s$ from 
these observables, the uncertainties should be attached to
both of these paramters which cannot be reduced within a purely
perturbative approach.
}

It may appear too pessimistic to estimate $\delta m$ from $\delta\bar V$.
One can argue that instead of considering the truncated asymptotic series 
for $\alpha_{_{\rm R}}$ the exact expression (\ref{richcoupling}) should 
be used in the Schr\"odinger equation for the $Q\bar{Q}$ system.
Then $m_Q$ can be determined from the mass of the $Q\bar{Q}$ bound-states
without any theoretical uncertainty.
This argument is based, however, on the complete knowledge of the function
$\alpha_{_{\rm R}}(q)$ including the range of its argument where
non-perturbative contributions are very important. In purely perturbative
calculations $\delta m$ has to be interpreted as a perturbative theoretical
uncertainty in $m_Q$ due to the asymptotic character of the series
(\ref{relrich}).

It seems reasonable to expect that the qualitative features of the
asymptotic expansion (\ref{result}) should be similar for the
Richardson potential and for the QCD potential.
This is a useful assumption because only limited information is
available in the latter case.
In perturbative QCD the coupling $\alpha_{_V}$ is expressed as an asymptotic
series in $\alpha_{\overline{MS}}$ and only the first three terms are 
known~\cite{fischler,mp}.
\begin{figure}[tb]\begin{center}
\vspace{2cm}
 \epsfxsize 12cm \mbox{\epsfbox{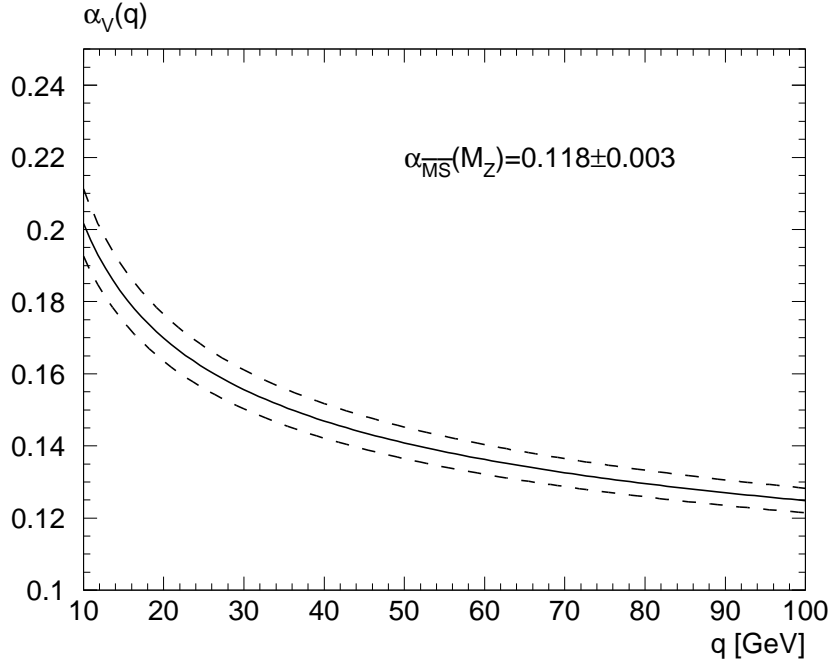}}
 \end{center}
 \caption[]{\label{fig2}
The momentum-space couplings $\alpha_{_V}^{(5)}(q)$ 
for different values of $\alpha_{\overline{MS}}(m_Z)$.
}
\end{figure}
In Fig.~\ref{fig2}, $\alpha_{_V}(q)$ is shown for $n_f=5$ active flavours and
$q$ between 10 and 100~GeV.
The solid line corresponds to $\alpha_{\overline{MS}}(m_Z)=0.118$ and
the dashed lines to changing this input value by $\pm 0.003$ which is the 
present error of this parameter~\cite{pdg}.
In QCD the first three coefficients of the $\beta$ function
(\ref{betafn}) are also known~\cite{QCDbeta,fischler,mp}:
\bea
\beta_{_{V,0}} &=& \beta_0 ,
\nonumber \\
\beta_{_{V,1}} &=& \beta_1 = 102 - {38\over 3}\,n_f ,
\nonumber \\
\beta_{_{V,2}} &=& 
  1854 - {2239\over 6}\, n_f + {377\over 54}\,{n_f}^2 \, + 
  \,{{\pi }^2}  \left( 18 - {{3\,{{\pi }^2}}\over 4} \right)\,
  \left( 33 - 2\,{n_f} \right) \,
\nonumber \\
  && + \left( 726 - {704\over 3}\,{n_f} + {104\over 9}\,{n_f}^2
       \right) \,{\zeta}(3)  .
\eea
Using this information one can write a formula for $\bar{\alpha}_{_V}(1/r)$
which is accurate up to fifth order in $\alpha_{_V}$:
\begin{eqnarray}
\bar\alpha_{_V} &=& \alpha_{_V} \,
\left\{ 1 \, + \,
  2\,{{{\beta_0}}^2}\,{f_2}\,{\rm a_v}^2 + 
   \left( 5\,{\beta_0}\,{\beta_1}\,{f_2} + 
      6\,{{{\beta_0}}^3}\,{f_3} \right) \,{\rm a_v}^3 + 
\right.
\nonumber \\  &&
\left.
   \left[ \left( 3\,{{{\beta_1}}^2} + 
6\,{\beta_0}\,{\beta_{V,2}} \right) \,
       {f_2} + 26\,{{{\beta_0}}^2}\,{\beta_1}\,{f_3} + 
      24\,{{{\beta_0}}^4}\,{f_4} \right] \,{\rm a_v}^4 
\right\} +  {{\cal O}(\alpha_{_V}^6)} ,
\end{eqnarray}
where $\alpha_{_V} \equiv \alpha_{_V}(q=1/r')$ and
${\rm a_v} = {\alpha_{_V}/( 4\pi)}$.
Let us remark that the term quadratic in $\alpha_{_V}$ is absent
because it has been absorbed by shifting the argument of the coupling
$\alpha_{_V}$.
After simple algebra the following formula is derived:
\begin{eqnarray}
\bar\alpha_{_V} &=&  \alpha_{_V}\left\{ \, 1 \, + \,
\left( 11 - \frac{2}{3}\, n_f \right)^2 \frac{\pi^2}{3} \,
\,{\rm a_v}^2 \, +  \right.
\nonumber\\  &&
\left[ \,
      \left( 935 - {1555\over 9} {n_f} + 
      {190\over 27} {n_f}^2 \right)\, \pi^2
      +  \left( 21296 - 
      3872 {n_f} + 
      {704 \over 3}{n_f}^2  - 
      {128\over 27} {n_f}^3 \right)\,{\zeta}(3) \, \right]\,
      {\rm a_v}^3 + 
\nonumber\\  &&
\left[ \, \left( 
      25596 - 
      {39797\over 6} n_f + 
      {21913\over 54} {n_f}^2 - 
      {377\over 81} {n_f}^3\,\right)\,\pi^2 +
      \left( 
      {376189\over 15} - 
      {237952\over 45} {n_f} + \right.\right.
\nonumber\\  && \left.\left.
      {19472\over 45}\,{n_f}^2 - 
      {6688\over 405}\,{n_f}^3 + 
      {304\over 1215}\,{n_f}^4\, \right)\,\pi^4 +
       \left( \,
      - {1089\over 4} + 33\,{n_f} - 
      {n_f}^2\,\right)\,\pi^6 \right.
\nonumber\\  && \left.  + \left(
      855712 - 
      {1889888\over 9}\,{n_f} + 
      {432640\over 27}\,{n_f}^2 - 
      {31616\over 81}\,{n_f}^3 \,\right) \,{\zeta}(3) \right.
\nonumber\\  && \left.\left.
      + \left(
      7986 - 
      {9196\over 3}\,{n_f} + 
      {2552\over 9}\,{n_f}^2 - 
      {208 \over 27}\,{n_f}^3 
     \,\right) \,\pi^2 \,{\zeta}(3)
       \right] 
    {\rm a_v}^4 + {\cal O}({\rm a_v}^5) \, \right\} .
\label{nfrel}
\end{eqnarray}
The numerical values of the coefficients in the above expansions
for $n_f=5$ are equal to:
\begin{eqnarray}
\bar\alpha_{_V} &=&  \alpha_{_V} \, + \,
 1.22454\,{{\alpha_{_V} }^3} \, +\,  5.59618\,{{\alpha_{_V} }^4} 
\, +\,  32.2015\,{{\alpha_{_V} }^5}\, +\,{\cal O}({\alpha_{_V} }^6) ,
\label{numrel1}
\end{eqnarray}
whereas for $n_f=4$ and $n_f=3$ one obtains
\begin{eqnarray}
\bar\alpha_{_V} &=&  \alpha_{_V} \, + \,
 1.44676\,{{\alpha_{_V} }^3} \, +\,  7.38182\,{{\alpha_{_V} }^4} 
\, +\,  46.3717\,{{\alpha_{_V} }^5}\, +\,{\cal O}({\alpha_{_V} }^6)
\label{numrel2}
\\
\bar\alpha_{_V} &=&  \alpha_{_V} \, + \,
 1.68750 \,{{\alpha_{_V} }^3} \, +\,  9.45282\,{{\alpha_{_V} }^4} 
\, +\,  64.0389\,{{\alpha_{_V} }^5}\, +\,{\cal O}({\alpha_{_V} }^6)
\label{numrel3}
\end{eqnarray}
respectively.
The numerical coefficients in the relations (\ref{numrel1})-(\ref{numrel3})
are large.
The asymptotic character of the expansion is evident and only for
very short distances ($1/r \simgt 30$~GeV) all five terms should be kept.
At such distances the number of active flavours is $n_f=5$.
The ${\cal O}(\alpha_{_V}^4)$ and ${\cal O}(\alpha_{_V}^5)$ terms 
in (\ref{numrel1}) are equal for 
$\alpha_{_V} = 0.1738$ and their contributions to $\bar{\alpha}_{_V}$ are
0.0051.
This implies that the truncation of the asymptotic series leads
to an uncertainty of about 2.8\% in $\bar{\alpha}_{_V}$.
The value of $\alpha_{_V}$ corresponds to $1/r = 32$~GeV, a distance which
is not very different from those probed in $t\bar{t}$ production
near the threshold.
Phenomenological consequences of the above observation are discussed
elsewhere~\cite{JKPST}.
At the distances probed by $b\bar{b}$ states, the corresponding values
of $\alpha_{_V}$ are so large that the quartic term in the expansion
(\ref{nfrel}) must be rejected.
Thus it would be simply inconsistent to keep other contributions
of this order like those in the relation between $\alpha_{_V}$ and
$\alpha_{\overline{MS}}$ even if this relation were known.
Our analysis leads to a rather surprising conclusion that in the
framework of purely perturbative QCD, theoretical uncertainties cannot be 
reduced below the present level !
(This is consistent with the discussion given in~\cite{PY}.)
If we accept such a radical point of view, we can proceed even further
by observing that the cubic term in (\ref{nfrel}) depends only on
$\beta_0$ and is the same for the QCD static potential and for the
Richardson potential.
The latter case suggests that it is not obvious at all if the inclusion of
the cubic term improves the accuracy of the formula for $\bar{\alpha}_{_V}$.
It may well be that a more precise answer is obtained if the cubic term
is also rejected.
Without extra information beyond the purely perturbative approach, we
cannot answer the question: to reject or not to reject.
Clearly, keeping the cubic term implies a stronger attraction between
the quarks.
Thus, if $m_b$ is determined from the energy of the $b\bar{b}$ ground state, 
the
value of $m_b$ obtained using a perturbative calculation and the potential
including the cubic term is larger than the value corresponding to 
rejecting this term.
The difference in $m_b$ is of the order of 100~MeV and our analysis
indicates that attempts to reduce this error to a much smaller value
within perturbative QCD may be inconsistent.

\section*{\bf Acknowledgements}
This work is partly supported by KBN grant 2P03B08414, by BMBF grant
POL-239-96 and by the Alexander von Humboldt Foundation.

\end{document}